\documentclass [prl,amsmath,showpacs,twocolumn,preprintnumbers,superscriptaddress]{revtex4-1}

\usepackage[latin1]{inputenc}
\usepackage{mathtools}
\usepackage{amssymb}
\usepackage{bm}
\usepackage{graphicx}

\RequirePackage{xspace}

\begin{document}

\title{Apparent Lorentz violation with  superluminal Majorana neutrinos at OPERA?}

\author{F. Tamburini}
\email{fabrizio.tamburini@unipd.it}
\affiliation{Department of Astronomy, University of Padova, vicolo dell' Osservatorio 3, Padova, Italy}

\author{M. Laveder}
\email{marco.laveder@unipd.it}
\affiliation{Department of Physics, University of Padova, via Marzolo 8, Padova, Italy and INFN - Sezione di Padova}

\begin{abstract}
From the data release of OPERA - CNGS experiment \cite{opera}, and publicly announced on 23 September 2011, we cast a phenomenological model based on a Majorana neutrino state carrying a fictitious imaginary mass term, already discussed by Majorana  in 1932. This imaginary mass term might be induced  in the Earth's crust during the 735 Km travel. Within the experimental errors, we prove that this hypothesis fits with OPERA and MINOS data and with the propagation of electron anti-neutrinos in the stellar structure of supernova SN1987a. Possible violations to Lorentz invariance due to quantum gravity effects have been considered.  
\end {abstract}


\maketitle

\section{Introduction}
Usually superluminal propagation and/or Lorentz invariance are associated with quantum gravity phenomenologies.
Quantum-gravitational (QG) effects  are thought  to generate  a violation of Lorentz invariance (LI) in the interactions of energetic neutrinos with the foamy structure of space-time, with the result that they do not propagate at the speed of light $c$ \cite{ellis2008probes,sakharov2009exploration}. 
High energy neutrinos are thought to be one of the best candidates to reveal QG phenomena \cite{PhysRevD.77.053014}.
Recent experimental results from the light analysis of the farthest quasars observed with the Hubble Space Telescope \cite{tamburini2011no} and from the propagation of gamma rays from INTEGRAL/IBIS observations of GRB041219A \cite{PhysRevD.83.121301}, demonstrate that the scales at which quantum gravity fluctuations are expected to occur and violate LI are much closer to the fundamental Planck's scale than that invoked in the literature to explain neutrino oscillations and propagation and the problem of sterile neutrino states.
Upper limits posed from the results of supernova SN1987a explosion \cite{hirata1987observation,cullen1999sn}, indicated that in the case of  linear violation of LI for energetic electron anti-neutrinos ($\bar \nu_e$) they are $2.7 \times 10^{10}$ GeV for subluminal propagation and  $2.5 \times 10^{10}$ for superluminal propagation. For a quadratic violation of the LI, instead, they are on the order of $4 \times 10^{4}$ GeV. The potential sensitivity to QG fluctuations expected from OPERA long-baseline neutrino experiment would be $4 \times 10^{8}$ GeV and  $7 \times 10^{5}$ GeV for the linear and quadratic LI.
The data released by OPERA suggest that the muon neutrino propagates at a speed larger than of light, $(v-c)/c = (2.48 \pm 0.28 \, (stat) \pm 0.30 \, (sys)) \times 10^{-5}$ \cite{opera}.
If these QG limits apply to neutrinos, those produced by OPERA would be free of these effects.
In OPERA experiment, the energetic muonic neutrinos ($\nu_\mu$), mainly produced in the decay, $\pi^{(\pm)} \rightarrow \mu^{(\pm)} +  \nu_\mu \;    (\bar \nu_\mu)$, cross the Earth's crust which is a structured dense medium with variable densities in space.
In this experiment, neutrinos seem to behave, during the $735$ km travel, either as tachyons or as they had a pseudo-tachyonic behavior when transversing a material, like photons in metamaterials \cite{pendry2006controlling,ziolkowski2001superluminal} with negative refractive index or less than unity, that allows an apparent superluminal propagation without violating causality.
The data released by CERN and Gran Sasso Labs are made of 16111 events detected in OPERA and correspond to about $10^{20}$ protons on target collected during the 2009, 2010 and 2011 CNGS runs. 
If they are free from systematic errors and with a correct interpretation, these data indicate that $\nu_\mu$'s could actually propagate with a superluminal speed after having crossed $735$ Km in the Earth's crust, with averaged density $\langle \rho \rangle_{\oplus} \simeq 2.7$~g/cm$^3$, in $2.5 \times 10^{-3}$ seconds. The effective resulting distance measured with GPS was $731278.0  \pm 0.2$ meters. The peak of neutrino detection occurs $\sim 6 \times 10^{-8}~$s before than expected, with a precision of $6.9 \, (stat) - 7.4 \, (sys)$ nanoseconds, which means $2.48 \times 10^{-5}$ times smaller than the light propagation time with $ 6 \sigma$ accuracy. The overall systematic uncertainty is obtained assuming independent error sources (see Tab. 2 of Ref. [1]).
A stringent limit of $|v-c| / c < 2 \times 10^{-9}$ was obtained with $\bar \nu_e$'s from the supernova (SN) explosion in the Magellanic Nebula SN1987a \cite{arnett1989supernova} after having traveled 51.4 kiloparsecs in vacuum and  $10^{12} - 10^{13}$ cm of dense stellar matter in expansion. This would indicate that either the neutrino is not a tachyon and OPERA assumptions are not correct, e.g. due to the orbital Earth motion that can give an apparent aberration of $\sim 6 \times10^{-8}$ s, or that the tachyonic effect could have occurred only inside the SN matter and that the propagation in space occurred at the speed of light $c$.

\section{Majorana Neutrino state}
Superluminal behaviors are mainly associated to tachyons, particles with imaginary mass. We apply Majorana theory \cite{Majorana:NC:1932,majorana1937theory} to build a phenomenological model of a Majorana neutrino with imaginary mass to explain OPERA anomaly. Within this hypothesis we show that neutrinos, when traversing layers of matter and/or interacting with sterile neutrinos inside gravitational fields, can behave like a tachyonic Majorana neutrino.

In his original work, Ettore Majorana formulated a particular solution of the Dirac equation with positive-defined mass solution,
\begin{equation}
\left[  \frac Wc + \alpha \cdot \mathbf{p} - \beta m c \right] \mathbf{\Psi}=0,
\label{dirac}
\end{equation}
where $\Psi$ is the wavefunction, $\alpha$ and $\beta$ are the Dirac matrices, $m$ is the mass of the particle, $\mathbf{p}$ the momentum and $W$ the energy \cite{Dirac01021928,thaller1992dirac}. From the relativistic formulation of the mass and energy, the energy of the particle is $W=\sqrt{c^2 p^2 + m^2 c^4}$ and the indetermination in the sign, interpreted by Dirac in terms of particle and antiparticle states, can be overcome by finding positive-definite rest mass solutions to the Dirac equation, for any spin value.
The solutions to Eq. \ref{dirac}, representing plane waves with positive-defined mass, are obtained from those derived through a relativistic transformation of zero momentum waves, whose energy is given by
\begin{equation}
W_0=\frac{m c^2}{ s+  1/2}
\label{maj1}
\end{equation}
and the intrinsic angular momentum parameter, $s$, describes the scalar, bosonic or fermionic solution. This equation supersedes the well-known Einstein's equation $E=mc^2$. 
Being for neutrinos $s= 1/2$, the two relations coincide. Moreover, in this case, CPT invariance, intimately related with Lorentz invariance (LI) violation, is not preserved being Majorana theory non-local \cite{Casalbuoni:2006fa}. 
Some hints of CPT violations have been observed with $\nu_\mu$'s \cite{PhysRevLett.107.021801} and with $\nu_e$'s  \cite{PhysRevD.82.113009}.
The immediate consequence to the choice of this set of solutions, derived from the set of infinitesimal Lorentz transformations, is that the spectrum of these particles exhibits a relationship between the intrinsic spin angular momentum and the Majorana-mass term, $m$, related to the particle's rest mass or to the acquired virtual mass.
Particles with different intrinsic angular momenta then present different masses, that are determined in the particle's reference frame through Eq. \ref{maj1}.
The relationship between spin and mass is valid for both bosonic ($s=1, \, 2, \, 3, \, ... \, $) and fermionic solutions ($s=1/2, \, 3/ 2, \, 5/2, \, ... \, $), and for  scalar particles, $s=0$,
\begin{equation} 
M =\frac {m}{s+1/2}.
\label{majmass}
\end{equation}
Eq. \ref{majmass} describes an infinite spectrum of particles with positive-defined mass values that decrease when the spin angular momentum of the particle increases. Photons in vacuum represent a particular solution of the bosonic case with zero rest mass. In addition to states with positive-definite mass values, there are other solutions found and discussed by Majorana himself, for which energy is related to momentum through
\begin{equation}
W= \pm \sqrt{c^2 p^2 - k^2 c^4}
\end{equation}
and they exist for all the positive values of $k$ for which $p \geq kc$ holds, derived directly from the elementary Lorentz transformation matrices, and that do not violate Lorentz invariance. Those states can be considered as belonging to the class of solution with imaginary mass term $\mathrm{i} k$ \cite{Majorana:NC:1932}.

\subsection{Tachyonic neutrino effects in dense media?}
From the results of OPERA and MINOS \cite{PhysRevD.76.072005}, we only make the \textit{ansatz} that the cause of a possible superluminal propagation of neutrinos might be due by their interaction inside matter and we verify this hypothesis by assuming that a similar behavior could have occurred to $\bar \nu_e$'s, independently from the neutrino flavor, only inside the matter of the exploding supernova SN1987a. If the $\bar \nu_e$'s follow the same distribution of OPERA and MINOS, according to the stellar structure parameters, they would confirm that it is a tachyonic propagation of a peculiar neutrino state inside Earth the possible cause of OPERA anomaly.
Calculations show that the additional contribution of the interstellar medium during the travel of the neutrino beam can be neglected.
According to a recent claim \cite{ceg} (thereafter, CG), no pseudo-superluminal motion in OPERA could have occurred with standard neutrinos because of the possible disruption of the beam shape due to effects induced by weak-current phenomena. An energetic standard neutrino, traveling faster than light in that medium, is expected to produce electron/anti-electron pairs that radiate away the neutrino energy. This radiation has not been observed neither with Opera \cite{opera} and Icarus \cite{icarus} nor from the moon as radio signals \cite{Stal&al:PRL:2007}. 
Thus, we agree with CG conclusions that Standard Model tachyonic neutrinos cannot be reconciled with the OPERA results. From this we suggest that Beyond-Standard Model Majorana neutrinos with imaginary mass $\mathrm{i} k$ obeying $p \geq kc$, may fit with Opera data, when crossing a medium.
In fact, in the case of a tachyonic behavior and standard dispersion relation, the pair production would require that $E_\nu^2-p_\nu^2 > (2m_e)^2$, not satisfied by an imaginary mass value. Consider a modified dispersion relation $E^2=p^2+m^2+F \ ,$
where $F$ is an arbitrary function. Here, $p$ should be considered as the conjugate momentum, so that, in principle, $p$ and $F$ may depend on space-time coordinates in a rather subtle way that may depend from the structure of space-time itself. It is clear that, depending on $p$ and $F$, this forbids the CG pair production.
The imaginary mass in this case is $\sqrt{m^2+F}$ with a negative $F$.
A possible explanation of this behavior can be due to a sterile neutrino mixing confined inside a region where a gravitational field is present or that the presence of matter/gravitational field introduces a preferred reference frame violating CPT symmetry and/or LI. Another cause could be the coupling of neutrinos with structured matter that can give rise to parametric resonances \cite{akhmedov2000parametric}, but all this goes far beyond the purposes of a phenomenological model presented in this letter.
The Majorana mass/spin relationship then becomes
\begin{equation}
m^2=-  k^2 \left(s + \frac 12 \right)^2 = - k^2.
\label{immass}
\end{equation}
When interacting with structured matter, neutrinos are expected also to acquire orbital angular momentum (OAM) and behave like a Majorana particle \cite{tamburini2011storming,mendonca2008neutrino,tamburini2010photon} obeying a more general mass/angular momentum relationship
\begin{equation}
M_\nu =  \frac{m}{\Sigma(\ell,q)+ 1/2 },
\label{procamajorana}
\end{equation}
where $\Sigma(\ell,q)$ is a general function of the spin $s=1/2$ and OAM $\ell$ of the photon and of the characteristic spatial scale of the perturbation $q$; the Majorana mass term coincides with that acquired by the neutrino in the unperturbed medium.
A similar effect is expected also when a beam of neutrinos crosses a Petrov-type D gravitational field like that of a rotating black hole \cite{tamburini2011twisting}.
Thus, differently from the space-time manifold structure of the Lorentz group, in which space is homogeneous and isotropic and time homogeneous, a medium can exhibit peculiar spatial structures that breaks the space-time symmetry.
In this case, for the peculiar solution with imaginary mass state, the mass/spin angular momentum relationship (\ref{immass}) is replaced by a mass/total angular momentum relationship $m^2= - k^2 \left( \Sigma(\ell,q) + \frac 12 \right)^2$.
This additional effect could give an additional hint to model the pulse and thus the time of arrival of neutrinos: the OAM-induced mass would act as a negative-squared mass term due to the inhomogeneities of the medium.

We now test our conjecture by comparing the results of OPERA, MINOS and those obtained from the propagation of the electronic anti-neutrinos only inside the matter of SN1987a. If these neutrinos  obey the Majorana condition, $p \geq kc$ and follow the same distribution of OPERA and MINOS, then the tachyonic behavior should have occurred only because of the interaction inside the SN matter and, in vacuum, neutrinos would propagate at a speed less or equal that of light, confirming our \textit{ansatz}. If these tachyonic properties of neutrinos were also present in vacuum or due to QG effects, we would have observed a neutrino peak years before the optical detection of the SN event, and this was not the case.
In Tab. 1 are reported the averaged neutrino energies for OPERA and MINOS and those calculated for SN1987a, the relative time anticipation $\Delta T / T_0$ and the imaginary mass terms, with uncertainties, obtained with Huzita relationship $m^2=2 E^2 \Delta T / T_0$ \cite{huzita1987neutrino}.

To calculate the neutrino mass in the SN medium we proceed with two independent approaches. In the first one, we calculated the neutrino propagation as it had a path length $l_{SN}$ inside a shell of matter with averaged density equivalent to that of the Earth's crust. The parameters of the SN here considered were taken from those of its precursor, Sanduleak -69$^\circ$ 202a, just before the SN explosion, a blue supergiant with radius $(1.8 - 4) \times 10^{12}$~cm and a mass of $3.96 \times 10^{34}$~g. More details can be found in Ref. \cite{petschek1990supernovae}. 
Assuming that there are no tachyonic effects in vacuum, then the time deviation calculated considering the propagation $l_{sp}$ in free space,

\begin{equation}
\left(\frac{\Delta T}{T_0} \right)_{SN} =  \frac { \langle \rho \rangle_{SN}} {\langle \rho \rangle_{\oplus}}    \; \frac{l_{sp}}{l_{SN}} \frac{|v-c|}{c}
\end{equation}
where $T_0$ is the photon flight-time needed to cross the SN distance, $\langle \rho \rangle_{SN} \sim (1.6 - 9.1) \times 10^{-3}$~g/cm$^{-3}$, the averaged densities of the SN, obtained by assuming a  $10^{12}$~cm and $1.8 \times 10^{12}$~cm radius, respectively. The values are reported in Tab. 1 as SN1987a-1 and SN1987a-2.
The second approach is based on the idea that the neutrino burst detected by Hirata et al. \cite{hirata1987observation} at $20$~MeV, indicating the formation of a neutron star, was actually affected by this pseudo-tachyonic behavior 
(SN1987a-3) and, surprisingly, coincides with the result of SN1987a-2.
\bigskip
\\
\begin{tabular}{|c|c|c|c|c|}
\hline
Dataset  &  $E (GeV) $    &   $\Delta T / T_0$  & $m (GeV) $   &$ \Delta m$ \\
\hline
OPERA  high & $42.9$ & $2.76 \times 10^{-5}$ & $0.317$  &  $0.093$\\
OPERA 17GeV  & $17$ & $2.48 \times 10^{-5}$ & $0.119$  &  $0.028$\\
OPERA  low   & $13.9$ & $2.18 \times 10^{-5}$ & $0.092$  &  $0.035$\\
MINOS        &  $3$ & $5.10  \times 10^{-5}$  & $0.030$  & $0.017$\\
SN1987a-1    & $ 2 \times 10^{-2}$ & $1.17  \times 10^{-1}$ & $0.009$ & $0.006$\\
SN1987a-2    & $ 2 \times 10^{-2}$ & $2.00  \times 10^{-2}$ & $0.004$ & $0.001$\\
SN1987a-3    & $ 2 \times 10^{-2}$ & $2.00  \times 10^{-2}$ & $0.004$ & $0.001$\\
\hline
\end{tabular}
\begin{figure}[!htb]
\centering
\includegraphics[width=8.8cm, keepaspectratio]{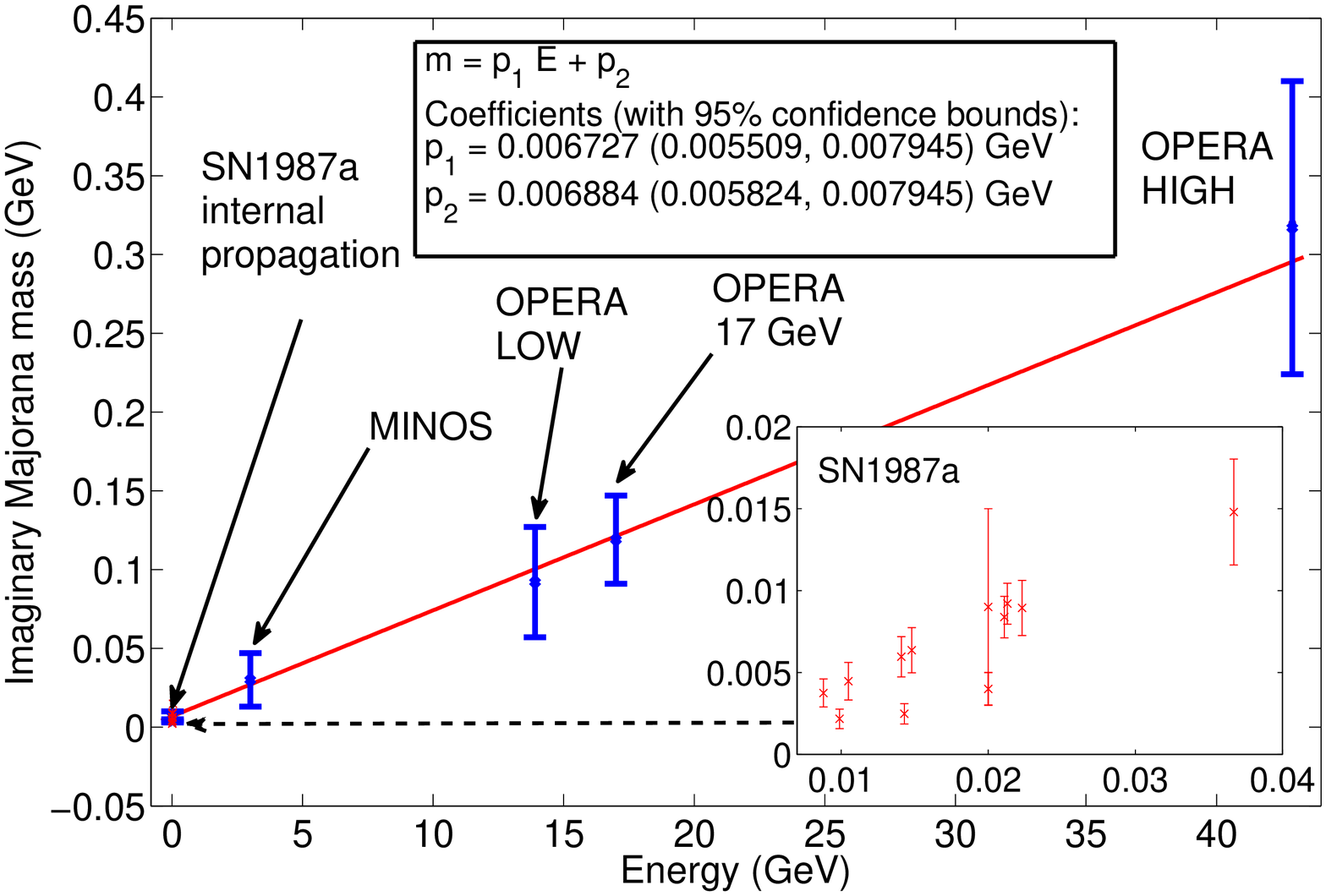}
\caption{Majorana fictitious imaginary mass vs. neutrino energy. The neutrinos propagating inside SN1987a and those from MINOS and OPERA data are distributed along a single line, indicating the presence of a variable imaginary mass term and the validity of the condition $p\geq kc$. Inset: zoom of dataset SN1987a-3.}
\label{fig1}
\end{figure}

All the energy and momentum  values follow a linear distribution, $m=p_1 E + p_2$, with 95\% confidence bounds. The fitting parameters and their intervals $p_1 =   0.006727 \;  (0.005509, 0.007945)$~GeV and $p_2 =   0.006884 \;  (0.005824, 0.007945)$~GeV, which is the tachyonic mass term of the neutrino in the limit $p=kc^2$. The sum of squares due to error, SSE=0.6483, and the root mean square error, RMSE=0.4026, indicate that the properties of $\bar \nu_e$'s inside the SN are in good agreement, with minor deviations, from OPERA and MINOS results. All obey the tachyonic relationship $p\geq kc$. Each single experimental event of SN1987a fit the family of distributions given by $p_1$ and $p_2$ values confirming our \textit{ansatz}.

\section{Conclusions}
The recent data released by OPERA experiment suggest a superluminal propagation of muonic neutrinos in the Earth's crust, if the data acquisition and interpretation are correct.
We propose a phenomenological description, based on Majorana theory to explain this anomaly without invoking quantum gravity effects.
These neutrinos are expected to behave like Majorana particles with an imaginary mass induced inside Earth's crust.  
The lacking of detection and energy loss from CG effect suggests the indirect evidence of a Majorana neutrino state with tachyonic behavior violating CPT invariance, obeying a non-local theory. This behavior could be due to MSW mixing, sterile neutrino states, parametric resonances and orbital angular momentum, but a clear explanation goes beyond the purposes of the present letter. We confirmed our hypothesis with the neutrinos of SN1987a, assuming an anomalous propagation only inside the supernova when it was starting its catastrophic collapse \cite{woosley1986physics}.
In vacuum, instead, neutrinos are expected to propagate at a speed less or equal that of light, otherwise the anticipation observed in the SN1987a would have been of years. Only future tests will contribute to solve the puzzling problem of neutrino propagation in matter.
Another possible explanation is that neutrinos have an apparent superluminal behavior similar to that of photons in a metamaterial. The group velocity, $v_g$, because of anomalous dispersion becomes apparently superluminal: during the propagation the incoming wave shape is distorted such that the amplitude at the propagating wave front increases while the amplitude in the tail decreases, often going below the detection limit. If the propagating wave position is measured at the field maximum it can be misinterpreted as a faster propagation of the propagating wave pulse as a whole with phase velocity, $v_p$  \cite{ziolkowski2001superluminal,smolyaninov2011metamaterial}. As $v_g>v_p$, the medium would cause the speed of only certain Fourier components of neutrino wavefunction in it to be larger than the speed of light in vacuum for a band of neutrino oscillation frequencies obeying the Majorana condition $p \geq kc$, but seems to be forbidden by CG effect for a standard neutrino. 
These neutrino properties, if confirmed, would not only give crucial information to the astrophysics of supernova explosions, compact objects, stellar interiors and cosmology \cite{1996ApJS}, but revolutionize the standard model of particles \cite{yao2006review} and its extensions to Lorentz-violating phenomenologies \cite{colladay1998lorentz}. If confirmed, in a more general scenario, Majorana theory can also predict the neutral Higgs boson mass: from Eq. \ref{majmass}, the mass of the spin $0$ particle is equal to $3$ times that of $Z^0$, i.e., $m=( 273.56 \pm 0.01 )$~GeV. 

\acknowledgments
The authors acknowledge Carlo Giunti, Antonio Masiero, Marco Matone, Massimo Della Valle and Cesare Chiosi  for the helpful discussions.

%


\begin{thebibliography}{10}%
\makeatletter
\providecommand \@ifxundefined [1]{%
 \ifx #1\undefined \expandafter \@firstoftwo
 \else \expandafter \@secondoftwo
\fi
}%
\providecommand \@ifnum [1]{%
 \ifnum #1\expandafter \@firstoftwo
 \else \expandafter \@secondoftwo
\fi
}%
\providecommand \enquote [1]{``#1''}%
\providecommand \bibnamefont  [1]{#1}%
\providecommand \bibfnamefont [1]{#1}%
\providecommand \citenamefont [1]{#1}%
\providecommand\href[0]{\@sanitize\@href}%
\providecommand\@href[1]{\endgroup\@@startlink{#1}\endgroup\@@href}%
\providecommand\@@href[1]{#1\@@endlink}%
\providecommand \@sanitize [0]{\begingroup\catcode`\&12\catcode`\#12\relax}%
\@ifxundefined \pdfoutput {\@firstoftwo}{%
 \@ifnum{\z@=\pdfoutput}{\@firstoftwo}{\@secondoftwo}%
}{%
 \providecommand\@@startlink[1]{\leavevmode}%
 \providecommand\@@endlink[0]{}%
}{%
 \providecommand\@@startlink[1]{%
  \leavevmode
  \pdfstartlink
   attr{/Border[0 0 1 ]/H/I/C[0 1 1]}%
   user{/Subtype/Link/A<</Type/Action/S/URI/URI(#1)>>}%
  \relax
 }%
 \providecommand\@@endlink[0]{\pdfendlink}%
}%
\providecommand \url  [0]{\begingroup\@sanitize \@url }%
\providecommand \@url [1]{\endgroup\@href {#1}{\urlprefix}}%
\providecommand \urlprefix [0]{URL }%
\providecommand \Eprint[0]{\href }%
\@ifxundefined \urlstyle {%
  \providecommand \doi [1]{doi:\discretionary{}{}{}#1}%
}{%
  \providecommand \doi [0]{doi:\discretionary{}{}{}\begingroup
  \urlstyle{rm}\Url }%
}%
\providecommand \doibase [0]{http://dx.doi.org/}%
\providecommand \Doi[1]{\href{\doibase#1}}%
\providecommand \bibAnnote [3]{%
  \BibitemShut{#1}%
  \begin{quotation}\noindent
    \textsc{Key:}\ #2\\\textsc{Annotation:}\ #3%
  \end{quotation}%
}%
\providecommand \bibAnnoteFile [2]{%
  \IfFileExists{#2}{\bibAnnote {#1} {#2} {\input{#2}}}{}%
}%
\providecommand \typeout [0]{\immediate \write \m@ne }%
\providecommand \selectlanguage [0]{\@gobble}%
\providecommand \bibinfo [0]{\@secondoftwo}%
\providecommand \bibfield [0]{\@secondoftwo}%
\providecommand \translation [1]{[#1]}%
\providecommand \BibitemOpen[0]{}%
\providecommand \bibitemStop [0]{}%
\providecommand \bibitemNoStop [0]{.\EOS\space}%
\providecommand \EOS [0]{\spacefactor3000\relax}%
\providecommand \BibitemShut [1]{\csname bibitem#1\endcsname}%
\bibitem{opera}%
  \BibitemOpen
  \bibfield{author}{%
  \bibinfo {author} {\bibnamefont{the OPERA~collaboration}},\ }%
  \bibinfo {journal} {Arxiv preprint 1109.4897v1}%
  \bibAnnoteFile{NoStop}{opera}%
\bibitem{ellis2008probes}%
  \BibitemOpen
\bibfield{journal}{%
    }%
  \bibfield{author}{%
  \bibinfo {author} {\bibfnamefont{J.}~\bibnamefont{Ellis}},  \emph{et~al.},\ }
  \bibfield{journal}{%
  \bibinfo {journal} {Phys. Rev. D}\ }%
  \textbf{\bibinfo {volume} {78}},\ \bibinfo {pages} {033013} (\bibinfo {year}
  {2008})%
  \bibAnnoteFile{NoStop}{ellis2008probes}%
\bibitem{sakharov2009exploration}%
  \BibitemOpen
  \bibfield{author}{%
  \bibinfo {author} {\bibfnamefont{A.}~\bibnamefont{Sakharov}},  \emph{et~al.},\ }
  in\ \emph{\bibinfo {booktitle} {Journal of Physics: Conference Series}},\
  Vol.\ \bibinfo {volume} {171}\ (\bibinfo {organization} {IOP Publishing},\
  \bibinfo {year} {2009})\ p.\ \bibinfo {pages} {012039}%
  \bibAnnoteFile{NoStop}{sakharov2009exploration}%
\bibitem{PhysRevD.77.053014}%
  \BibitemOpen
\bibfield{journal}{%
    }%
  \bibfield{author}{%
  \bibinfo {author} {\bibfnamefont{N.~E.}\ \bibnamefont{Mavromatos}}
  \emph{et~al.},\ }%
  \bibfield{journal}{%
  \bibinfo {journal} {Phys. Rev. D}\ }%
  \textbf{\bibinfo {volume} {77}},\ \bibinfo {pages} {053014} (\bibinfo {year}
  {2008})%
  \bibAnnoteFile{NoStop}{PhysRevD.77.053014}%
\bibitem{tamburini2011no}%
  \BibitemOpen
  \bibfield{author}{%
  \bibinfo {author} {\bibnamefont{{Tamburini, F.}}}, \bibinfo {author}
  {\bibnamefont{{Cuofano, C.}}}, \bibinfo {author} {\bibnamefont{{Della Valle,
  M.}}},\ and\ \bibinfo {author} {\bibnamefont{{Gilmozzi, R.}}},\ }%
  \bibfield{journal}{%
  \Doi{10.1051/0004-6361/201015808}{\bibinfo {journal} {A\&A}}\ }%
  \textbf{\bibinfo {volume} {533}},\ \bibinfo {pages} {A71} (\bibinfo {year}
  {2011})%
  \bibAnnoteFile{NoStop}{tamburini2011no}%
\bibitem{PhysRevD.83.121301}%
  \BibitemOpen
  \bibfield{author}{%
  \bibinfo {author} {\bibfnamefont{P.}~\bibnamefont{Laurent}} \emph{et~al.},\
  }%
  \bibfield{journal}{%
  \bibinfo {journal} {Phys. Rev. D}\ }%
  \textbf{\bibinfo {volume} {83}},\ \bibinfo {pages} {121301} (\bibinfo {month}
  {Jun}\ \bibinfo {year} {2011})%
  \bibAnnoteFile{NoStop}{PhysRevD.83.121301}%
\bibitem{hirata1987observation}%
  \BibitemOpen
  \bibfield{author}{%
  \bibinfo {author} {\bibfnamefont{K.}~\bibnamefont{Hirata}} \emph{et~al.},\ }%
  \bibfield{journal}{%
  \bibinfo {journal} {Phys. Rev. Lett.}\ }%
  \textbf{\bibinfo {volume} {58}},\ \bibinfo {pages} {1490} (\bibinfo {year}
  {1987})%
  \bibAnnoteFile{NoStop}{hirata1987observation}%
\bibitem{cullen1999sn}%
  \BibitemOpen
  \bibfield{author}{%
  \bibinfo {author} {\bibfnamefont{S.}~\bibnamefont{Cullen}}\ and\ \bibinfo
  {author} {\bibfnamefont{M.}~\bibnamefont{Perelstein}},\ }%
  \bibfield{journal}{%
  \bibinfo {journal} {Phys. Rev. Lett.}\ }%
  \textbf{\bibinfo {volume} {83}},\ \bibinfo {pages} {268} (\bibinfo {year}
  {1999})%
  \bibAnnoteFile{NoStop}{cullen1999sn}%
\bibitem{pendry2006controlling}%
  \BibitemOpen
  \bibfield{author}{%
  \bibinfo {author} {\bibfnamefont{J.}~\bibnamefont{Pendry}}, \bibinfo {author}
  {\bibfnamefont{D.}~\bibnamefont{Schurig}},\ and\ \bibinfo {author}
  {\bibfnamefont{D.}~\bibnamefont{Smith}},\ }%
  \bibfield{journal}{%
  \bibinfo {journal} {Science}\ }%
  \textbf{\bibinfo {volume} {312}},\ \bibinfo {pages} {1780} (\bibinfo {year}
  {2006})%
  \bibAnnoteFile{NoStop}{pendry2006controlling}%
\bibitem{ziolkowski2001superluminal}%
  \BibitemOpen
  \bibfield{author}{%
  \bibinfo {author} {\bibfnamefont{R.}~\bibnamefont{Ziolkowski}},\ }%
  \bibfield{journal}{%
  \bibinfo {journal} {Phys. Rev. E}\ }%
  \textbf{\bibinfo {volume} {63}},\ \bibinfo {pages} {046604} (\bibinfo {year}
  {2001})%
  \bibAnnoteFile{NoStop}{ziolkowski2001superluminal}%
\bibitem{arnett1989supernova}%
  \BibitemOpen
  \bibfield{author}{%
  \bibinfo {author} {\bibfnamefont{W.}~\bibnamefont{Arnett}}, \bibinfo {author}
  {\bibfnamefont{J.}~\bibnamefont{Bahcall}}, \bibinfo {author}
  {\bibfnamefont{R.}~\bibnamefont{Kirshner}},\ and\ \bibinfo {author}
  {\bibfnamefont{S.}~\bibnamefont{Woosley}},\ }%
  \bibfield{journal}{%
  \bibinfo {journal} {Ann. rev. of Astron. and Astroph.}\ }%
  \textbf{\bibinfo {volume} {27}},\ \bibinfo {pages} {629} (\bibinfo {year}
  {1989})%
  \bibAnnoteFile{NoStop}{arnett1989supernova}%
\bibitem{Majorana:NC:1932}%
  \BibitemOpen
  \bibfield{author}{%
  \bibinfo {author} {\bibfnamefont{E.}~\bibnamefont{Majorana}},\ }%
  \bibfield{journal}{%
  \bibinfo {journal} {Nuov.\ Cim.}\ }%
  \textbf{\bibinfo {volume} {9}},\ \bibinfo {pages} {335} (\bibinfo {year}
  {1932})%
  \bibAnnoteFile{NoStop}{Majorana:NC:1932}%
\bibitem{majorana1937theory}%
  \BibitemOpen
  \bibfield{author}{%
  \bibinfo {author} {\bibfnamefont{E.}~\bibnamefont{Majorana}},\ }%
  \bibfield{journal}{%
  \bibinfo {journal} {Nuov.\ Cim.}\ }%
  \textbf{\bibinfo {volume} {14}},\ \bibinfo {pages} {171} (\bibinfo {year}
  {1937})%
  \bibAnnoteFile{NoStop}{majorana1937theory}%
\bibitem{Dirac01021928}%
  \BibitemOpen
\bibfield{journal}{%
    }%
  \bibfield{author}{%
  \bibinfo {author} {\bibfnamefont{P.A.M.}\ \bibnamefont{Dirac}},\ }%
  \bibfield{journal}{%
  \bibinfo {journal} {Proc. R. Soc. Lon. A}\ }%
  \textbf{\bibinfo {volume} {117}},\ \bibinfo {pages} {610} (\bibinfo {year}
  {1928})%
  \bibAnnoteFile{NoStop}{Dirac01021928}%
\bibitem{thaller1992dirac}%
  \BibitemOpen
  \bibfield{author}{%
  \bibinfo {author} {\bibfnamefont{B.}~\bibnamefont{Thaller}},\ }%
  \emph{\bibinfo {title} {The Dirac Equation,}}\ \bibinfo {publisher} {Springer} \ (\bibinfo {year} {1992})%
  \bibAnnoteFile{NoStop}{thaller1992dirac}%
 \bibitem{Casalbuoni:2006fa}
   R.~Casalbuoni,
   PoS {\bf EMC2006} 004, Arxiv preprint hep-th/0610252 (2006).
%
\bibitem{PhysRevLett.107.021801}%
  \BibitemOpen
  \bibfield{author}{%
  \bibinfo {author} {\bibfnamefont{P.}~\bibnamefont{Adamson}} \emph{et~al.},\
  }%
  \bibfield{journal}{%
  \bibinfo {journal} {Phys. Rev. Lett.}\ }%
  \textbf{\bibinfo {volume} {107}},\ \bibinfo {pages} {021801} (\bibinfo {year}
  {2011})%
  \bibAnnoteFile{NoStop}{PhysRevLett.107.021801}%
\bibitem{PhysRevD.82.113009}%
  \BibitemOpen
  \bibfield{author}{%
  \bibinfo {author} {\bibfnamefont{C.}~\bibnamefont{Giunti}}\ and\ \bibinfo
  {author} {\bibfnamefont{M.}~\bibnamefont{Laveder}},\ }%
  \bibfield{journal}{%
  \Doi{10.1103/PhysRevD.82.113009}{\bibinfo {journal} {Phys. Rev. D}}\ }%
  \textbf{\bibinfo {volume} {82}},\ \bibinfo {pages} {113009} (\bibinfo {year}
  {2010})%
  \bibAnnoteFile{NoStop}{PhysRevD.82.113009}%
\bibitem{PhysRevD.76.072005}%
  \BibitemOpen
  \bibfield{author}{%
  \bibinfo {author} {\bibfnamefont{P.}~\bibnamefont{Adamson}} \emph{et~al.},\
  }%
  \bibfield{journal}{%
  \bibinfo {journal} {Phys. Rev. D}\ }%
  \textbf{\bibinfo {volume} {76}},\ \bibinfo {pages} {072005} (\bibinfo {year}
  {2007})%
  \bibAnnoteFile{NoStop}{PhysRevD.76.072005}%
\bibitem{ceg}%
  \BibitemOpen
  \bibfield{author}{%
  \bibinfo {author} {\bibfnamefont{A.G.}~\bibnamefont{Cohen}},\ }%
  {\bibfnamefont{S.L.}~\bibnamefont{Glashow}}, \bibinfo {author}
  \bibinfo {journal} {Arxiv preprint 1109.6562v1}%
  \bibAnnoteFile{NoStop}{ceg}%
\bibitem{icarus}%
  \BibitemOpen
  \bibfield{author}{%
  \bibinfo {author} {\bibnamefont{the ICARUS~collaboration}},\ }%
  \bibinfo {journal} {Arxiv preprint 1110.3763v1}%
  \bibAnnoteFile{NoStop}{icarus}%
\bibitem{Stal&al:PRL:2007}%
  \BibitemOpen
  \bibfield{author}{%
  \bibinfo {author} {\bibfnamefont{O.}~\bibnamefont{St{\aa}l}}  \emph{et~al.},\
  }%
  \bibfield{journal}{%
  \bibinfo {journal} {Phys.\ Rev.\ Lett.}\ }%
  \textbf{\bibinfo {volume} {98}},\ \bibinfo {pages} {071103} (\bibinfo {year} {2007})%
  \bibAnnoteFile{NoStop}{Stal&al:PRL:2007}%
\bibitem{akhmedov2000parametric}%
  \BibitemOpen
  \bibfield{author}{%
  \bibinfo {author} {\bibfnamefont{E.}~\bibnamefont{Akhmedov}},\ }%
  \bibfield{journal}{%
  \bibinfo {journal} {Pramana}\ }%
  \textbf{\bibinfo {volume} {54}},\ \bibinfo {pages} {47} (\bibinfo {year}
  {2000})%
  \bibAnnoteFile{NoStop}{akhmedov2000parametric}%
\bibitem{tamburini2011storming}%
  \BibitemOpen
  \bibfield{author}{%
  \bibinfo {author} {\bibfnamefont{F.}~\bibnamefont{Tamburini}}\ and\ \bibinfo
  {author} {\bibfnamefont{B.}~\bibnamefont{Thid{\'e}}},\ }%
  \bibinfo {journal} {Arxiv preprint 1105.0700}%
  \bibAnnoteFile{NoStop}{tamburini2011storming}%
\bibitem{mendonca2008neutrino}%
  \BibitemOpen
\bibfield{journal}{%
    }%
  \bibfield{author}{%
  \bibinfo {author} {\bibfnamefont{J.}~\bibnamefont{Mendon{\c{c}}a}}\ and\
  \bibinfo {author} {\bibfnamefont{B.}~\bibnamefont{Thid{\'e}}},\ }%
  \bibfield{journal}{%
  \bibinfo {journal} {EPL}\ }%
  \textbf{\bibinfo {volume} {84}},\ \bibinfo {pages} {41001} (\bibinfo {year}
  {2008})%
  \bibAnnoteFile{NoStop}{mendonca2008neutrino}%
\bibitem{tamburini2010photon}%
  \BibitemOpen
  \bibfield{author}{%
  \bibinfo {author} {\bibfnamefont{F.}~\bibnamefont{Tamburini}}, \bibinfo
  {author} {\bibfnamefont{A.}~\bibnamefont{Sponselli}}, \bibinfo {author}
  {\bibfnamefont{B.}~\bibnamefont{Thid{\'e}}},\ and\ \bibinfo {author}
  {\bibfnamefont{J.}~\bibnamefont{Mendon{\c{c}}a}},\ }%
  \bibfield{journal}{%
  \bibinfo {journal} {EPL (Europhysics Letters)}\ }%
  \textbf{\bibinfo {volume} {90}},\ \bibinfo {pages} {45001} (\bibinfo {year}
  {2010})%
  \bibAnnoteFile{NoStop}{tamburini2010photon}%
\bibitem{tamburini2011twisting}%
  \BibitemOpen
  \bibfield{author}{%
  \bibinfo {author} {\bibfnamefont{F.}~\bibnamefont{Tamburini}}, \bibinfo
  {author} {\bibfnamefont{B.}~\bibnamefont{Thid\'e}}, \bibinfo {author}
  {\bibfnamefont{G.}~\bibnamefont{Molina-Terriza}},\ and\ \bibinfo {author}
  {\bibfnamefont{G.}~\bibnamefont{Anzolin}},\ }%
  \bibfield{journal}{%
  \bibinfo {journal} {Nature Physics}\ }%
  \textbf{\bibinfo {volume} {7}},\ \bibinfo {pages} {195} (\bibinfo {year}
  {2011})%
  \bibAnnoteFile{NoStop}{tamburini2011twisting}%
\bibitem{akhmedov2001floquet}%
  \BibitemOpen
  \bibfield{author}{%
  \bibinfo {author} {\bibfnamefont{E.}~\bibnamefont{Akhmedov}},\ }%
  \bibfield{journal}{%
  \bibinfo {journal} {Physics of Atomic Nuclei}\ }%
  \textbf{\bibinfo {volume} {64}},\ \bibinfo {pages} {787} (\bibinfo {year}
  {2001})%
  \bibAnnoteFile{NoStop}{akhmedov2001floquet}%
\bibitem{huzita1987neutrino}%
  \BibitemOpen
  \bibfield{author}{%
  \bibinfo {author} {\bibfnamefont{H.}~\bibnamefont{Huzita}},\ }%
  \bibfield{journal}{%
  \bibinfo {journal} {Modern Physics Letters A}\ }%
  \textbf{\bibinfo {volume} {2}},\ \bibinfo {pages} {905} (\bibinfo {year}
  {1987})%
  \bibAnnoteFile{NoStop}{huzita1987neutrino}%
\bibitem{petschek1990supernovae}%
  \BibitemOpen
  \emph{\bibinfo {title} {Supernovae}},\ edited by\ \bibinfo {editor}
  {\bibfnamefont{A.}~\bibnamefont{Petschek}}\ (\bibinfo {publisher} {New York,
  NY (USA); Springer-Verlag New York Inc.},\ \bibinfo {year} {1990})%
  \bibAnnoteFile{NoStop}{petschek1990supernovae}%
\bibitem{woosley1986physics}%
  \BibitemOpen
  \bibfield{author}{%
  \bibinfo {author} {\bibfnamefont{S.}~\bibnamefont{Woosley}}\ and\ \bibinfo
  {author} {\bibfnamefont{T.}~\bibnamefont{Weaver}},\ }%
  \bibfield{journal}{%
  \bibinfo {journal} {Ann. rev. of astron. and astroph.}\ }%
  \textbf{\bibinfo {volume} {24}},\ \bibinfo {pages} {205} (\bibinfo {year}
  {1986})%
  \bibAnnoteFile{NoStop}{woosley1986physics}%
\bibitem{smolyaninov2011metamaterial}%
  \BibitemOpen
  \bibfield{author}{%
  \bibinfo {author} {\bibfnamefont{I.}~\bibnamefont{Smolyaninov}},\ }%
  \bibfield{journal}{%
  \bibinfo {journal} {Journal of Optics}\ }%
  \textbf{\bibinfo {volume} {13}},\ \bibinfo {pages} {024004} (\bibinfo {year}
  {2011})%
  \bibAnnoteFile{NoStop}{smolyaninov2011metamaterial}%
\bibitem{1996ApJS}%
  \BibitemOpen
  \bibfield{author}{%
  \bibinfo {author} {\bibfnamefont{N.}~\bibnamefont{{Itoh}}}, \bibinfo {author}
  {\bibfnamefont{H.}~\bibnamefont{{Hayashi}}}, \bibinfo {author}
  {\bibfnamefont{A.}~\bibnamefont{{Nishikawa}}},\ and\ \bibinfo {author}
  {\bibfnamefont{Y.}~\bibnamefont{{Kohyama}}},\ }%
  \bibfield{journal}{%
  \Doi{10.1086/192264}{\bibinfo {journal} {ApJS}}\ }%
  \textbf{\bibinfo {volume} {102}},\ \bibinfo {pages} {411} (\bibinfo {year}
  {1996})%
  \bibAnnoteFile{NoStop}{1996ApJS}%
\bibitem{yao2006review}%
  \BibitemOpen
  \bibfield{author}{%
  \bibinfo {author} {\bibfnamefont{W.}~\bibnamefont{Yao}} \emph{et~al.},\ }%
  \bibfield{journal}{%
  \bibinfo {journal} {Journal of Physics G: Nuclear and Particle Physics}\ }%
  \textbf{\bibinfo {volume} {33}},\ \bibinfo {pages} {1} (\bibinfo {year}
  {2006})%
  \bibAnnoteFile{NoStop}{yao2006review}%
\bibitem{colladay1998lorentz}%
  \BibitemOpen
  \bibfield{author}{%
  \bibinfo {author} {\bibfnamefont{D.}~\bibnamefont{Colladay}}\ and\ \bibinfo
  {author} {\bibfnamefont{V.}~\bibnamefont{Kosteleck{\`y}}},\ }%
  \bibfield{journal}{%
  \bibinfo {journal} {Phys. Rev. D}\ }%
  \textbf{\bibinfo {volume} {58}},\ \bibinfo {pages} {116002} (\bibinfo {year}
  {1998})%
  \bibAnnoteFile{NoStop}{colladay1998lorentz}%
 \end{thebibliography}

\end{document}